# SuperCell: A Wide-Area Coverage Solution Using High-Gain, High-Order Sectorized Antennas on Tall Towers

Pratheep Bondalapati*, Abhishek Tiwari, Mustafa Emin Sahin, Qi Tang, Srishti Saraswat, Vishvas Suryakumar, Ali Yazdan, Julius Kusuma and Amit Dubey

*Abstract*—In this article we introduce a novel solution called SuperCell, which can improve the return on investment (ROI) for rural area network coverage. SuperCell offers two key technical features: it uses tall towers with high-gain antennas for wide coverage and high-order sectorization for high capacity. We show that a solution encompassing a high-elevation platform in excess of 200 meters increases coverage by 5x. Combined with dense frequency reuse by using as many as 36 azimuthal sectors from a single location, our solution can adequately serve the rural coverage and capacity demands. We validate this through propagation analysis, modeling, and experiments.

The article gives a design perspective using different classes of antennas: Luneburg lens, active/passive phased array, and spatial multiplexing solutions. For each class, the corresponding analytical model of the resulting signal-to-interference plus noise ratio (SINR) based range and capacity prediction is presented. The spatial multiplexing solution is also validated through field measurements and additional 3D ray-tracing simulation. Finally, in this article we also shed light on two recent SuperCell field trials performed using a Luneburg lens antenna system. The trials took place in rural New Mexico and Mississippi. In the trials, we quantified the coverage and capacity of SuperCell in barren land and in a densely forested location, respectively. In the article, we demonstrate the results obtained in the trials and share the lessons learned regarding green-field and brown-field deployments.

## I. INTRODUCTION

The wireless industry develops new technologies to drive advancements in cellular standards. However, according to the most recent data from the GSMA [1], [2], there are still 600 million people globally living outside of areas covered by mobile broadband networks. This lack of coverage is particularly concentrated in rural and remote areas, especially in regions like Sub-Saharan Africa, which is home to 67% of the world's uncovered population. While the number of people online has been increasing every year, the GSMA says only about 40 percent of people living in lower-middle-income countries are connected, compared with nearly 75% of the population in high-income countries.

In urban solutions, the focus is on capacity enhancements using high-end technologies such as massive MIMO, multisector antennas, and cell densification through the use of macro/small cells. For rural connectivity, however, there is significantly less technology investment since the economics are not attractive. In such scenarios, a large number of such macrocells have to be deployed for adequate coverage, each serving a limited number of users. Considering that the macrocells also need to be supplied with power and backhaul connectivity, it is not hard to imagine why the business case does not close.

For rural connectivity in developing countries, lack of coverage is a key challenge. Further, in many areas that lack connectivity, the population is distributed sparsely and the average revenue per user (ARPU) is low. In particular, the cost to acquire and build a cell site is very high and constitutes a significant portion of the total cost.

The conventional way of providing cellular coverage is to use 4G macro-cell sites. In challenging rural areas, this means that many such macro-cell sites have to be built to provide coverage, or conversely, each of the macro-cells covers only a few users. Either way, the business case can fail as cost-tocoverage is poor. At the same time, we need a solution that provides sufficient capacity per user.

To address the challenge of cost-to-coverage while maintaining high capacity, we propose SuperCell, a novel connectivity solution optimized for rural areas. SuperCell comprises two primary elements: 1) a high-elevation platform with highgain antennas to obtain a wide coverage, 2) high-order sectorization to provide high capacity through heavy frequency reuse. In this article we propose an empirical solution that is field-validated for practical feasibility, to encourage future exploration and optimization.

To enable faster time-to-market, we focus on using commercial off-the-shelf (COTS) components as much as possible. The use of standards-based cellular (2G, 3G or 4G) technology is therefore a hard requirement as it allows us to provide service without proprietary radio systems and user equipment, and leverage economies of scale. In other words, such a system does not require specialized UEs or any design changes to the older generation UEs. SuperCell presents a novel development path focused on coverage for rural connectivity, as shown in Fig. 1.

To build such a system, we first consider the appropriate antenna systems to provide coverage and high-order sectorization (HOS) for spatial multiplexing. Therefore, we seek directional antennas with high gain. We consider their performance, and their suitability for commercial deployment on high-elevation platforms. Next, we consider signal propagation, in particular we focus on (1) path loss & coverage prediction, and (2)

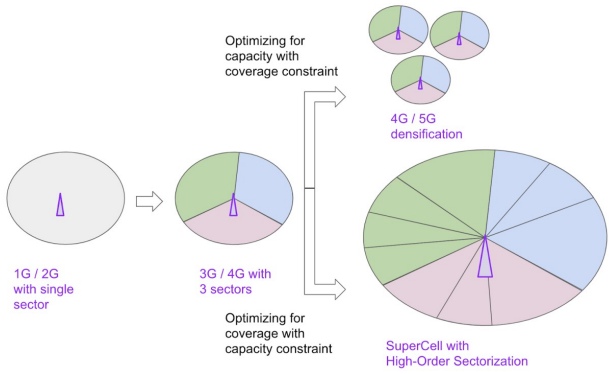

Fig. 1: Proposed SuperCell coverage vs. conventional 1G/2G, 3G/4G coverage

azimuthal spreading due to the environment.

To build such a system, we first consider the appropriate antenna systems to provide coverage and high-order sectorization (HOS) for spatial multiplexing. Therefore, we seek directional antennas with high gain. We consider their performance, and their suitability for commercial deployment on high-elevation platforms. Next, we consider signal propagation, in particular we focus on (1) path loss & coverage prediction, and (2) azimuthal spreading due to the environment.

To validate our design and modeling results, we performed several SuperCell field trials using a Luneburg lens antenna system in various parts of rural US that represent the geographical conditions of potential deployment areas in the developing world. Two of such trials took place in rural New Mexico (NM) and Mississippi (MS).

## II. DESIGN CONSIDERATIONS FOR SUPERCELL

To derive our system design requirements, consider the LTE link budget shown in Table I. Transmit power in the uplink direction has more stringent limit than in the downlink direction. In the uplink, user devices / equipment (uEs) are limited to 23 dBm of power (and 26 dBm in Cat 2 uEs). To improve both uplink and downlink link budgets and increase coverage range, we focus on enhancing the antenna gain on the BTS side. The corresponding increase in coverage and capacity needs to be studied more carefully, in particular relative to the Maximum Allowable Path Loss (MAPL) to enable minimum SNR levels for LTE connectivity. MAPL is used to derive coverage range and area based on Standard Propagation Model (SPM) and/or field measurements, therefore it is an important metric for SuperCell.

In a typical LTE or a GSM system, the BTS antenna gain is approximately 18 dBi, with corresponding MAPL of 145 to 150 dB. Based on the SPMs tuned using the measurement results discussed in Section VI, a 250m tower hosting 18 dBi antenna should provide coverage up to 22km. This is only

2.75x improvement from a traditional macrocell radius (i.e. 8km). An economic viability analysis (not covered in the scope of this paper) shows a need for 5x improvement in cell radii. Assuming a pathloss exponent of 2, this requires a 28 dB improvement in link budget. Based on COTS equipment spec, we can add another 10 dB in the antenna gain. The remaining 18 dB can be achieved by increasing the tower height. It is well studied in text books [42] that, by assuming circularly symmetric structure, the beamwidth of the antenna is dependent on the maximum gain by the following equation.

| Parameter | Value (downlink) | Value (uplink) |
|---|---|---|
| Tx power | 46 dBm | 23 dBm |
| Tx ant gain | 28 dBi | 0 dBi |
| Pathloss | 170 dB | 170 dB |
| Rx ant gain | 0 dBi | 28 dBi |
| Rx power | -96 dBm | -119 dBm |
| Bandwidth | 20 MHz | 500 KHz |
| Noise power | -101 dBm | -117 dBm |
| Noise figure | 8 dB | 4 dB |
| SNR | -3 dB | - 6 dB |

TABLE I: Sample cellular communication link budget

$$BW = \sqrt{\frac{\mathcal{X}\eta}{G_{max}}} \qquad (1)$$

where $X$ and are specified a given type of antenna. A 28 dBi gain antenna can have a 3-dB beamwidth of only 6 degrees (both horizontal and vertical) at 2500 MHz, whereas the conventional BTS antenna with 18 dBi gain can have a 3-dB beamwidth of 50 degrees. With this beamwidth, each antenna of a conventional BTS can cover 120 degrees of azimuth, while higher-gain antenna can only cover an azimuth of 10 degree with 10-dB crossover points.

In summary, there are three challenges and the corresponding proposed solutions are enlisted below:

1) Coverage modeling We need accurate models of the pathloss for cellular deployments and planning. Unfortunately, there is little prior work relevant to SuperCell, therefore in Sec. VI we developed a field measurement system and calibrated a Standard Propagation Model (SPM).

2) Capacity enhancement Larger coverage area means higher capacity demand within one SuperCell BTS, and demand increases in quadratic scale with cell radius. We developed a high-order sectorization system to increase spatial efficiency, and in Sec. VIII we examine how propagation and azimuthal spreading affects capacity.

3) Antenna Systems Engineering SuperCell is much taller than conventional BTS. Its height is similar to conventional broadcast systems, however SuperCell has to enable two-way communication under permissible EIRP limits. Employing high-gain antennas can overcome this challenge. However, high-gain antennas tend to have larger surface areas that contribute to the wind loading effect on the tower. In Sec. IV we study how directional antennas can be built and mounted reliably on a tower.

## III. CHANNEL MODELS FOR SUPERCELL

Pathloss or propagation loss is a widely used term in the literature to quantify the loss in signal power from a Tx to Rx. The total power loss between a Tx-Rx, excluding cable losses and antenna gains, can include numerous factors such as free space pathloss, multipath fading, diffraction loss from terrain and clutter, shadow fading, penetration loss, atmospheric loss, rain loss.

TABLE II: Parameter description for SPM

| Parameter | Description |
| --- | --- |
| $K$ | multiplying factor for Log(d) constant offset (dB) |
| $d$ | distance between the receiver and the transmitter (m) |
| $H_{tx}$ | effective height of the transmitter antenna (m) |
| $K_4$ | multiplying factor for diffraction calculation |
| $K_5$ | multiplying factor for Log($H_{tx}$) x Log($d$)) |
| $K_6$ | multiplying factor for $H_{rx}$ |
| $K_7$ | multiplying factor for Log($H_{rx}$)) |
| $H_{rx}$ | mobile antenna height (m) |
| $K_{clutt}$ | multiplying factor for f(clutter)) |
| $f_{clutt}$ | average of weighted losses due to clutter |
| $L_{Diff}$ | losses due to diffraction over an obstructed path (dB) ) |
| $K_{hill}$ | LoS corrective factor for hilly regions |

Most cellular coverage models emphasize the first factors: free space and multipath losses. The remaining terms, such as shadowing, penetration, body and other loses are accounted as fading margin [16]. Free space pathloss is the only known term that can be calculated in closed-form using the Friis free space equation [15]. The two-ray model is often used to model the effect of multipath fading. There are several models in the literature to estimate diffraction loss. Both multipath fading and diffraction loss heavily depend on the terrain and clutter information at a given location. Another factor contributing to the observed pathloss for a given Tx-Rx location is the line of sight (LOS) condition. LOS is defined as visual (optical) line of sight in this work.

For ease of reference, the SPM equation is reproduced in 2 and the parameters are described in Table II.

$$PL(d,H_{tx},H_{rx}) = K_1 + K_2\text{Log}(d) + K_3\text{Log}(H_{tx})$$

$$+ K_4 L_{\text{Diff}} + K_5\text{Log}(d)Log(H_{tx}) +$$

$$K_6 H_{rx} + K_7\text{Log}(H_{rx})$$

$$+ K_{clutt}f_{clutt} + K_{hill,LOS}$$
(2)

To our best knowledge, there are no existing studies on rural macro-cell pathloss model that can accurately predict the path loss at large ranges i.e. greater than 20km for tower heights greater than 150m. Even in the 3GPP models, there is limitation on the maximum usable distance (known as breakpoint distance) and the tower height of the path loss model.

Small-scale fading occurs in wireless communication channel due to the existence of multipath. The multipath profile is dependent on the Tx-Rx distance, the environment in the vicinity of the Tx-Rx and random frequency modulation due to mobility of the Tx or Rx. The multipath effect can further be categorized as time dispersion and angle dispersion. One of the well established metrics to characterize the angular dispersion is known as angular spread (spatial domain equivalent to delay spread in time domain), defined as the standard deviation of power received in various azimuthal or elevation angles (discussed in detail in Section VIII). Once again, the studies done in the literature are limited to 70m tower heights and cell radii less than 2km [39].

## IV. CANDIDATE CLASS OF ANTENNA SYSTEMS FOR SUPERCELL

As a part of our study, we reviewed several commercial classes of antenna solutions (as possible empirical solution candidates) and identified the following to be suitable for SuperCell requirements: Luneberg Lens and Panel antennas. Note that this requirement engineering is applicable to any class of antennas as long as the given techno-economic cost model is satisfied.

### A. Luneberg Lens

The Luneburg lens is traditionally associated with the prospects of transforming a plane wave into a point-like spherical wave [24]. If a Luneburg lens, being passive, is excited by a plane wave from a feed, the wave front will be concentrated into a point, acting like an antenna array with relatively high gain and high side lobe rejection. Additionally, every point on the surface of an ideal Luneburg lens is the focal point of a plane wave incident from the other side, providing it 360 degree scan angle. Being highly directive, Luneburg lenses (LL) are useful for application in wireless communication, radio astronomy, radar, electronic warfare [25], [29].

*1) Design strategies:* A gradient of the refractive index generates a gradient index lens, where the refractive gradient can be spherical, axial or radial. Such gradual variation can be used to produce lenses without aberrations leading to focused beams. Luneburg lens antennas [26] are gradient index devices where the permittivity gradually varies from $\epsilon_r$ = 2, at the sphere center, to $\epsilon_r = 1$ (air) at the surface of the sphere. Every point on the surface of an ideal Luneburg lens is a focal point for a plane wave incident from the opposite side. The permittivity $\epsilon_r$ variation of the lens is given by

$$\epsilon_r(r) = 2 - \left(\frac{r}{R}\right)^2 \tag{3}$$

where $R$ is the radius of the lens [24] and r is the distance from the point to the center of the sphere. Fig. ?? shows the optical path of waves propagating into the lens. The permittivity distribution is extremely complicated to be precisely realized.

*2) Example design for SuperCell:* For the design of a wideband frequency of operation from 1795 MHz to 2600 MHz, a 10 diameter lens with 1.251 m outer diameter is chosen. The lens is realized with 6 concentric shells with the same width. The lens is fed using a pyramidal horn antenna whose phase center is aligned with the focal point of the lens for maximum gain achievement. The phase center of an antenna is an imaginary point from where the far field seems to radiate. The phase center of the horn antenna is located at its aperture.

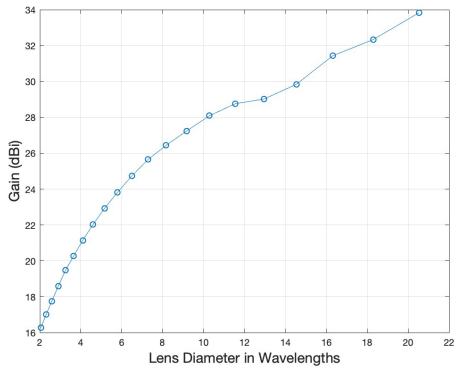

Fig. 2: Variation of gain as a function of lens diameter

The lens diameter is within the EPA constraints as with 1.251 m lens diameter the surface area can be calculated as 52.92 sq. ft, which is less than the effective projected area of 90 sq-ft. The lens diameter is also within the shipping currents., which limits antennas beyond 1.8 m. The gain of the simulated lens antenna is 28.8 dBi at 2600 MHz and 26.31 dBi at 1800 MHz. This is consistent with the fact that gain increases with frequency. The side lobe level achieved at 2600 MHz is 23.85 dB and at 1800 MHz is 18.13 dB below the main beam. A parametric study is carried out to understand the intricacies of the effect of diameter and the number of shells on the lens performance. The bandwidth of the lens is limited at the low frequency by the size of the lens. At higher frequencies, the lens performance is limited due to the decrease in the aperture efficiency. Additionally, the feed antenna pattern also limits the lens bandwidth. Additionally, as discussed previously, the increase in lens diameter improves the gain. Moreover, for a fixed lens diameter, increasing the number of shells improves the gain until a threshold is reached as shown in Fig. 2.

### B. Panel antennas for SuperCell

In the previous sections, although, the lens based solutions are proven to be a candidate class of antennas for SuperCell, there are certain drawbacks. A flexible non-uniform sectorization is difficult to achieve using the lens. On the other hand, panel antennas (antenna array systems) offer more flexibility in terms of non-uniform sectorization and avoiding cusping loss. Reprogramming the beam forming coefficients, in case of active systems, offers additional benefits of modifying the radiation patterns post-production and even post-deployment. Although, there are several realizations of panel antennas in the literature such as linear, circular, planar etc, the scope of this paper is to focus on the linear arrays.

### V. EPA Constraint for SuperCell Antennas
### A. Lens Antennas

For the Luneberg lens, the arrangement consists of spherical antennas symmetrically placed around the tower mast as shown in Fig. 14. In this case, the wind load scales based on antenna diameter. The nominal drag coefficient for a sphere is 0.5 and accordingly the EPA for the three spheres is given by:

$$\text{EPA}_L = 4.71 D^2 \tag{4}$$

where $D$ is the relative lens diameter expressed in terms of wavelengths. Note that the estimate is approximate since it does not include the contributions of reduced dynamic pressure for the spheres situated in the wake as well as not accounting for interference drag effects.

### B. Flat-Panel Arrays

The effective projected area of the SuperCell base station antenna is computed for two configurations: flat-panel arrays and Luneberg Lens. EPA estimates derived provide a wind loading design constraint towards selecting the optimal antenna choice.

Two flat-panel array configurations are considered: horizontal and vertically placed antennas installed around the tower mast. The horizontal antenna is assumed to be a short cylinder of radius, $R$, and length, $L$, with either side open. For $R \, L$, circumferential elements may be treated as flat plates to determine wind loads. Accordingly, the drag force acting on an infinitesimal element may be derived from [38]. The EPA is given by:

$$\text{EPA}_H = 2.35 N_v N_h \sqrt{\_\blacklozenge} \tag{5}$$

where $N_v$, $N_h$, and are the number of vertical and horizontal elements, and wavelength, respectively. For a vertical antenna, the cylinder is assumed to be long ($L \gg R$), and therefore, the flow around the cylinder may be assumed to be twodimensional. Assuming a drag coefficient of $1.2$, the EPA is given as :

$$\text{EPA}_V = 1.2 N_v N_h \sqrt{\_\blacklozenge}^2 \qquad (6)$$

For the same $N_v$, $N_h$ and , we see that horizontally-placed antennas incur almost twice more wind loading.

### C. EPA vs Antenna Gain

The design of tall towers is particularly sensitive to the wind load induced by the antennas much more than their weight loading contribution. This is due to the relatively higher wind speeds at the top of the tower and the large moments induced at the bottom of the tower due to the high wind loads. To compare the wind load performance of the three candidate antenna choices, the same antenna gain is used as baseline. The antenna gain for the flat plane array antennas is given as follows:

$$G = E_g + 10 \log_{10} N_h N_v \qquad (7)$$

where the element gain, $E_g$ varies with wavelength. For 2.5 GHz, an element of gain of 5 dBi maybe assumed. For the Luneberg Lens, the relationship of the gain vs relative antenna diameter is illustrated as follows:

| Gain | Vertical Placement | | Horizontal Placement | | Luneberg Lens | |
|---|---|---|---|---|---|---|
| (dBi) | Nv, Nh | EPA (ft²) | Nv, Nh | EPA (ft²) | D | EPA (ft²) |
| 30 | 56, 6 | 62 | 6, 56 | 122 | 15 | 158 |
| 28 | 42, 4 | 31 | 4, 42 | 61 | 10 | 66 |
| 25 | 32, 3 | 18 | 3, 32 | 35 | 7 | 33 |
| 23 | 24, 2 | 9 | 2, 24 | 17 | 5 | 17 |
| 20 | 18, 2 | 7 | 2,18 | 13 | 4 | 9 |

TABLE III: Variation of number of antenna elements for panel antennas for horizontal vs. vertical placement, calculated for 2500 MHz

Using the same antenna gain and assuming the same carrier frequency of 2500 MHz, the dimensions of the panel arrays and the lens antenna may be computed from Eq. 7 and Fig. 2 respectively from which the corresponding EPA may be computed. This is shown in Table. III. We note that for an equivalent gain, the vertically-placed antennas incur the lowest wind load. The horizontally-placed antennas and the Luneberg Lens incur similar, higher wind load.

### VI. Pathloss measurement campaign

We conducted the pathloss measurements in several locations in United States that represent the terrain types found in rural Africa such as flat crop land (Quad city, Illinois), hilly terrain (Frenchberg, Kentucky) and flat land with trees (Metcalf, Georgia). For all these locations, the tests were done at four tower heights: 30m, 60m, 120m and 250m. Conventional macro-cells' tower height fall under 30m-60m range, for which the pathloss models already exist and therefore serves as a baseline to compare the results at 2-4x taller scenarios. The subsections below describe these tests and discuss the resulting pathloss models:

### A. Quad city, Illinois

The tower is 300m tall, owned by American Tower Corporation (ATC) for TV broadcasting. The drive routes chosen for making the measurements up to 40km radius. The resulting heatmaps generated using the tuned models and a 36-sector Luneberg Lens antenna system is shown in Fig. 3. Note that a comparison to a conventional MC with height 30m and 3sector antenna system is also shown. Table IV lists the ratio of SC coverage radius to MC coverage radius as function of varying Reference signal receive power (RSRP) threshold levels. The resulting tuned model is shown in Fig. 6 for 2500 MHz (LTE cellular Band 41). The tuned SPM parameters based on measurement data is shown in Table. V.

### B. Metcalf, Georgia

The measurement campaign was extended to a tower location at Metcalf, Georgia. The tower is 320m tall, owned by ATC for TV broadcasting. The drive routes chosen for making the measurements up to 40km radius. The resulting heatmaps generated using the tuned models and a 36-sector Luneberg lens antenna system for SC and a 3-sector antenna system for conventional MC is shown in Fig. 4. Table VI lists the ratio of SC coverage radius to MC coverage radius as function of varying RSRP threshold levels.

| RSRP threshold (dBm) | SC ($km^2$) | MC ($km^2$) | SC/MC (ratio) |
|---|---|---|---|
| -120 | 3,853 | 160.84 | 24 |
| -115 | 2,557 | 101.66 | 25 |
| -110 | 1,699 | 68.27 | 25 |
| -105 | 842 | 34.89 | 24 |
| -100 | 536 | 23.52 | 23 |
| -95 | 230 | 12.15 | 19 |
| -90 | 144 | 7.79 | 18 |
| -85 | 57 | 3.42 | 17 |
| -80 | 29 | 2.04 | 14 |

TABLE IV: Variation of SC/MC coverage ratio as a function of required RSRP threshold at Quad city, IL

| K parameter | for 728 MHz | for 2500 MHz |
|---|---|---|
| $K_{1,los}$ | 6.35 | 4.89 |
| $K_{2,los}$ | 37.13 | 32.4 |
| $K_{1,nlos}$ | 0 | 0 |
| $K_{2,nlos}$ | 32.87 | 33.67 |
| $K_3$ | 8.2 | - 9.02 |
| $K_4$ | 0.48 | 0.4 |
| $K_5$ | -3.88 | 0 |
| $K_6$ | -0.12 | -0.09 |
| $K_7$ | -1.18 | -1.14 |

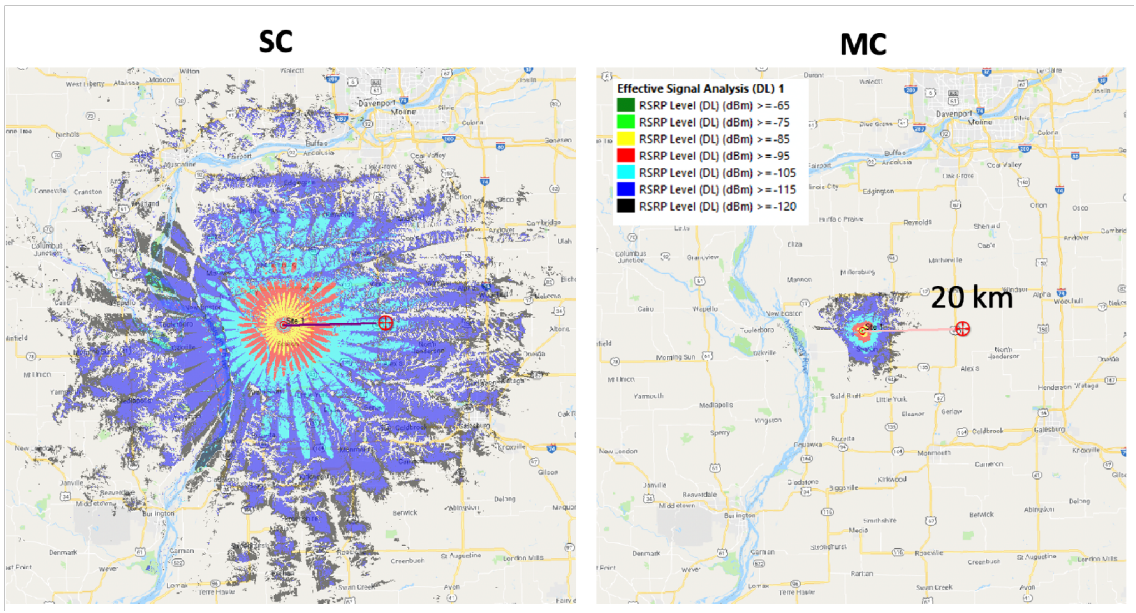

Fig. 3: RSRP heatmap for SC and MC at Band 41 in Quad city, IL

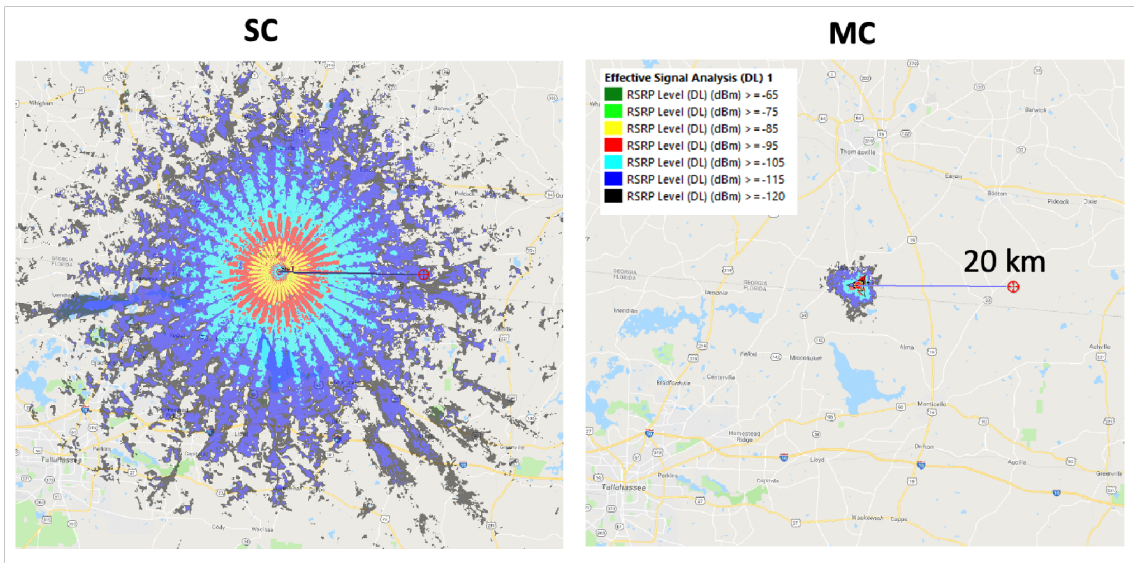

Fig. 4: RSRP heatmap for SC and MC at Band 41 in Metcalf, GA

| $L_{Dif}$ | 0 | 0 |
|---|---|---|
| $K_c$ | 1 | 1 |
| $K_{hill}$ | 0 | 0 |
| $f_c$ | 3 | 3 |

TABLE V: Resulting parameters for the tuned SPM for Quad city

| RSRP threshold (dBm) | SC ($km^2$) | MC ($km^2$) | SC/MC (ratio) |
|---|---|---|---|
| -120 | 2,044 | 30.96 | 66 |
| -115 | 1,241 | 19.11 | 65 |
| -110 | 847.7 | 13.55 | 63 |
| -105 | 454.3 | 7.99 | 57 |
| -100 | 295.2 | 5.47 | 54 |
| -95 | 136.1 | 2.96 | 46 |
| -90 | 84.9 | 1.77 | 48 |
| -85 | 33.6 | 0.58 | 58 |
| -80 | 16.9 | 0.32 | 53 |

TABLE VI: Variation of SC/MC coverage ratio as a function of required RSRP threshold at Metcalf, GA

| RSRP threshold (dBm) | SC ($km^2$) | MC ($km^2$) | SC/MC (ratio) |
|---|---|---|---|
| -120 | 544.1 | 10.79 | 50 |
| -115 | 280.5 | 8.51 | 33 |
| -110 | 147.6 | 6.29 | 23 |
| -105 | 88.5 | 3.62 | 25 |
| -100 | 58.3 | 2.44 | 24 |
| -95 | 35.6 | 0.97 | 37 |
| -90 | 22 | 0.49 | 45 |
| -85 | 14.6 | 0.24 | 61 |
| -80 | 8.8 | 0.15 | 59 |

TABLE VII: Variation of SC/MC coverage ratio as a function of required RSRP threshold at Frenchberg, KY

*C. Frenchberg, Kentucky*

The resulting heatmaps generated using the tuned models and a 36-sector Luneberg Lens antenna system for SC and a 3-sector antenna system for conventional MC is shown in Fig. 4. Table VII lists the ratio of SC coverage radius to MC coverage radius as function of varying RSRP threshold levels. It can be observed that unlike, Quad city and Metcalf, the SC coverage for Frenchberg is not uniform. This lack of coverage is due to the irregular terrain found in this area that is blocking the LOS viewshed from the tower.

In summary, the pathloss measurements carried out across a variety of terrain types indicate that there could be a significant improvement in the coverage area with increasing BTS heights in cellular communication systems. For example, it can be inferred from Fig. 6 that $160$ dB of pathloss is observed at $23$km for a $30$m BTS height, whereas similar values of pathloss is observed at $52$km for a $250$m BTS height. Note that the taller the tower, the more the coverage relative to a conventional Macrocell.

VII. RAY TRACING SIMULATION: PATHLOSS CALIBRATION

We carried out a study using 3D ray-tracing simulations to characterize propagation characteristics for low and high bands ($728$ MHz and $2500$ MHz), between a tall tower ($30$m-$250$m) and omni-directional user equipment. The study objectives include determining the impact of changing height of the tower-mounted antenna on the pathloss measurements using simulations and comparison of the results with the drive test campaign. The simulations were set up and executed using Remcom's Wireless InSite solution providing accurate, physics-based simulations with 30 cm clutter resolution.

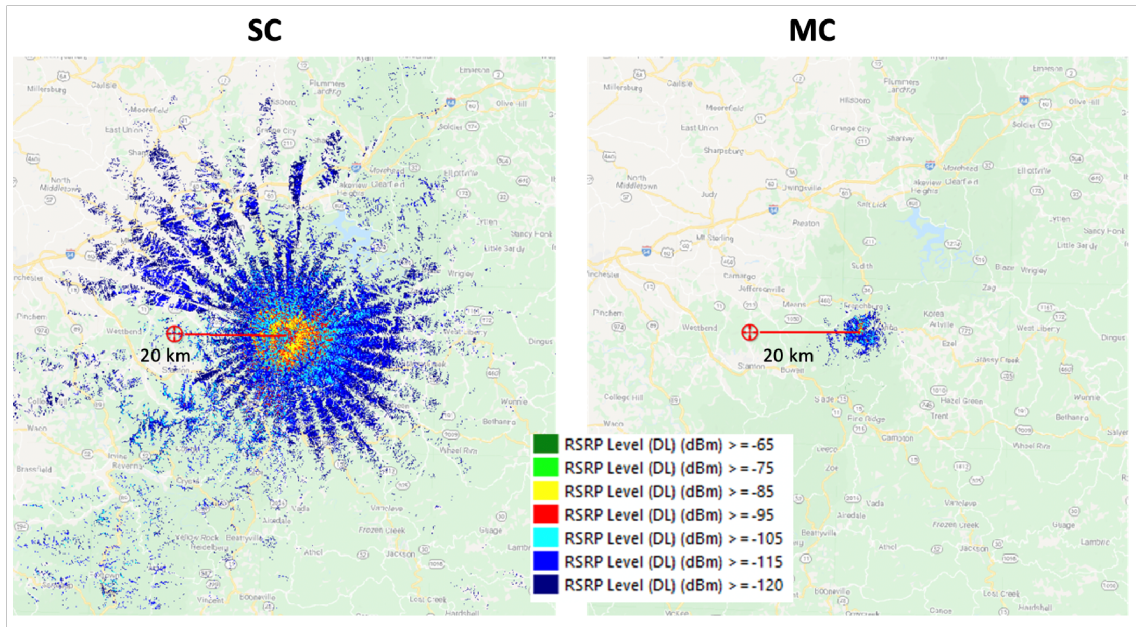

Fig. 5: RSRP heatmap for SC and MC at Band 41 in Frenchberg, KY

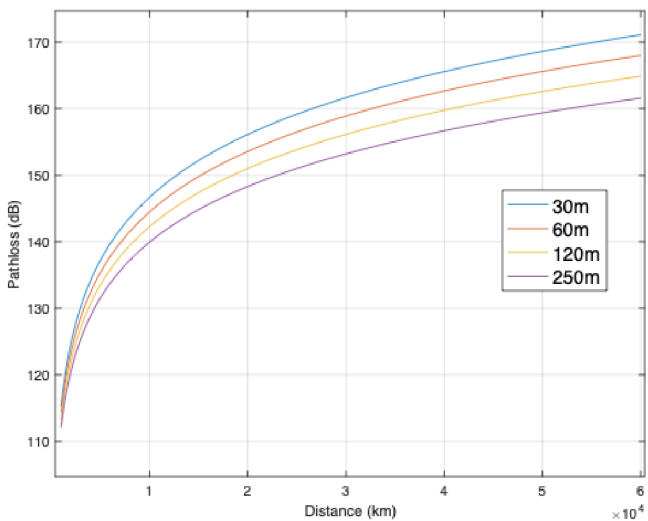

Fig. 6: Resulting pathloss model (SPM) for high-band (2500 MHz) in Quad city

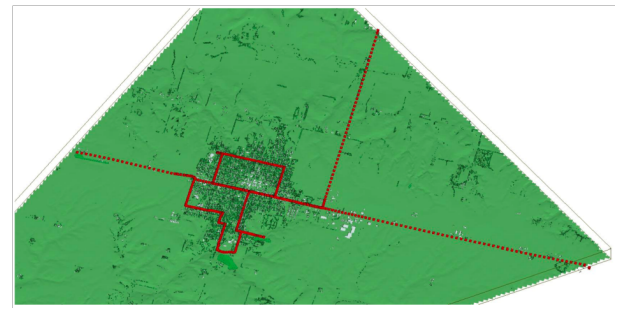

Fig. 7: Simulated Drive Test Points as Viewed within Wireless InSite

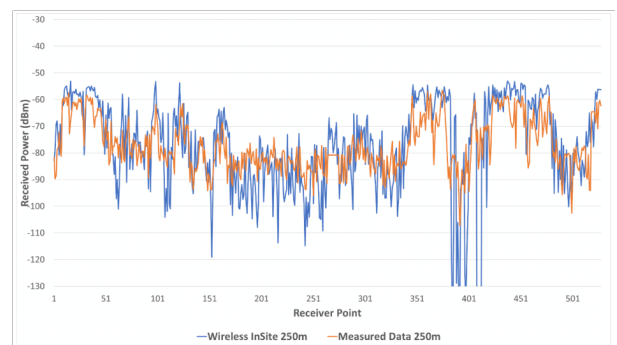

Fig. 8: Calibrated received powerQuad city comparison (mean error reduced to 0.72 dB); Orange: simulation, Blue: measured

*A. Pathloss simulation conditions*

There were originally approximately 78,000 measurement points; these points were down-sampled these to 528 points, to minimize the computational complexity of the 3D simulations. The final set of receiver points are shown from within Wireless InSite in Fig. 7. Next we performed an initial set of simulations, and used information provided by the measurements for the transmitter antenna and theoretical assumptions for the receiver antenna in order to calibrate results further.

From these results it is clear that the simulations are capturing the general trends in the received power along the route quite well, and that after scaling to account for actual antenna gains (the calibrated set in Fig. 8), the magnitudes are in reasonably good agreement.

| Error | Before | After |

| Statistics | calibration | calibration |
|---|---|---|
| Mean (dB) | 2.47 | 0.724 |
| Standard deviation (dB) | 11.88 | 11.87 |
| RMS (dB) | 12.13 | 11.90 |

TABLE VIII: Simulation vs. drive test error metrics

### B. Summary of PL simulation results

Table VIII provides several error metrics for the comparisons between the two data sets, before and after the calibration adjustments. As shown, with these minor adjustments, the mean error is reduced from approximately 2.5 dB to approximately 0.72dB. The mean of the absolute value of the error, the standard deviation, and the RMS error are reduced as well.

The comparisons were performed using the nearest neighbors to each of the simulated points. These points were generally separated by 15 to 20 meters or more. Given the amount of multipath and fading that might be expected, particularly within the town, it is likely that these points were under-sampling the variations along the route, which will result in increases to error metrics. A more involved study could evaluate the effect of fading on pathloss in more detail as described by $M_{ref,dB}$ in (13). This would be a detailed research topic in our future work.

### VIII. PAS MEASUREMENT CAMPAIGN

A simple way to measure PAS is using an adaptive steerable antenna that can be steered in azimuth axis. Such experiments have been done in the past for small range of coverage area [39]. However, making PAS measurements up to $2600$ MHz and a distance scale of $10$s of kilometers has several experiment design constraints. The antenna beamwidth requirement for a 1-degree resolution demands an antenna array of dimensions in the order of a few meters. There are no commercially available antenna array products that are capable of conducting this experiment. We propose a novel way to conduct this measurement using a narrow beam commercially available dish antenna.

A detailed mathematical analysis for Power azimuth spectrum (PAS) is provided in Appendix A, which also derives an overall capacity evaluation model for SuperCell. The PAS tests were conducted in two locations in the United States, Amarillo, Texas and Truth or Consequences, New Mexico.

### A. Amarillo, Texas

This location was primarily used to characterize the variation of PAS with respect to tower height. The tower rented for this test was approximately 300m tall. In order to be consistent with pathloss measurements, the PAS measurement tests were conducted at the same steps of height; 30m, 60m, 120m and 250m.

The PAS estimations made using the measured data at various locations are shown in Fig. 9 as a function of height. At 2.2km Tx-Rx distance, there is not a significant change in azimuth spread primarily due to LoS path being dominant.

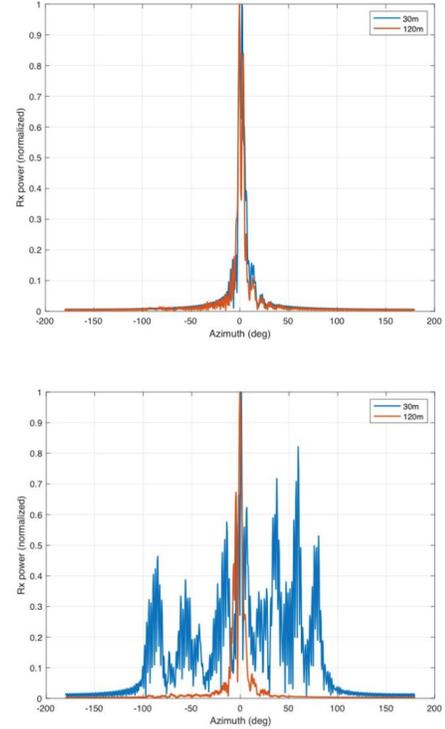

Fig. 9: Variation of PAS with elevation, measured in Amarillo

However, at 4km Tx-Rx distance, there is a significant reduction in azimuth spread when the tower height is increased to 120m.Fig 9 shows the variation of PAS as a function of these tower height combinations. It can be observed that with progressive increase in heights for a given distance, there is relative reduction in azimuth spread.

### B. Truth or Consequences, New Mexico

The reason for selecting this location is it is mountainous and can provide a perspective on the multipath reflections due to the hilly terrain. There is no clutter or trees in this location but a minor amount of tree bushes were scattered across the hilly surfaces. The tower was built on top of a mountain that is 310m above ground level (AGL). The antenna on the tower was mounted only at one height with bore-sight facing the east. The vehicle locations were chosen based on the road access convenience at roughly 3.5km, 16km, 20km and 35km.

### C. Principle component analysis

Based on the PAS measurement results discussed above, it can be observed that PAS could be a formulated as a function of various factors (or dimensions) such as elevation difference between the Tx-Rx, distance, delay spread and pathloss. Therefore PAS can be formulated as a multi-dimensional problem. Principle component analysis (PCA) tools can then be applied to evaluate the dependence of PAS on each of these dimensions. The best practice for this type of dimension reduction problem is to use Singular Value Decomposition (SVD) [46]. It can be concluded that the second dimension i.e. pathloss, has the highest dependence of 36.16%. Fig. 10 shows the variation of PAS as a function of pathloss and compares with the theoretical model. The relationship is found to obey an exponential curve fit.

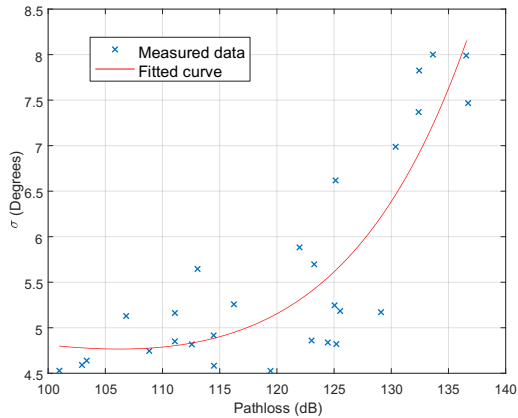

Fig. 10: Variation of PAS as a function of pathloss and a possible exponential curve fit

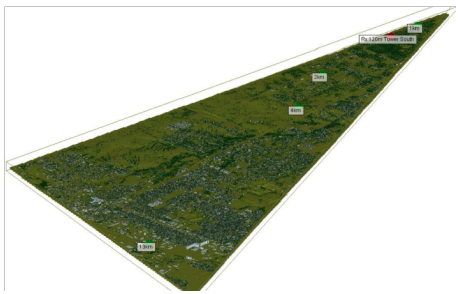

Fig. 11: Boundary of assessment around Amarillo, TX, viewed from within the Wireless InSite Graphical User Interface.

## IX. Ray tracing simulation: PAS analysis

This study was carried out to simulate PAS at three test measurement points (in alignment with a measurement campaign) and a grid of hypothetical vehicle locations within a predefined area covering urban, suburban, and rural areas to the North of downtown Amarillo, Texas. The study consisted of simulations to characterize propagation characteristics for high and low bands as well as angular spread analysis using 3D ray-tracing simulations . The study objectives included determining the impact of changing height of the BTS, placed on a tall tower(30m-250m), as well as the position of drive test vehicle within the different environments.

### A. PAS simulation setup

The area to the North of downtown Amarillo, Texas, and extends over urban, suburban, and rural areas of the city, with the Southern part of the area filled with a number of structures for homes and businesses. The scope of the study included simulation of propagation from four test measurement points to antennas placed on a tower at two different heights, 30 meters and 120 meters. The area of interest and the positions of the measurement points and tower are shown within Google

| Height (m) | (deg) |
|---|---|
| 30 | 0.35 |
| 60 | 0.31 |
| 120 | 0.2 |
| 250 | 0.1 |

TABLE IX: Variation of standard deviation of    with tower height

Earth in, below. The equivalent setup is shown within Wireless InSite in Fig. 11.

For the second part of the assessment, simulations were performed to predict coverage over the full area shown in Fig.11, sampled uniformly in space. This study used the same tower heights and antenna characteristics for the transmitters and receivers, but the tower-mounted antenna was specified as the transmitter to allow generation of a coverage map to the receivers in the scene. Approximately 500 receivers were placed uniformly throughout the triangular scene to provide general estimates of coverage. In addition to the initial specifications, the predictions also included a set of data using isotropic antennas at each end in order to collect statistics on propagation paths and uplink and downlink angles of arrival, independent of the antenna gain patterns.

### B. Summary of PAS simulation results

PAS field measurements were done first and used to calibrate the Wirless Insite simulation model. The calibrated model was then used to extrapolate the PAS measurements over a wide coverage area. This was repeated for all four tower heights: 30m, 60m,120m and 250m. The resulting standard deviation of    as a function of tower height is summarized in Table IX.

Now that we have a consensus on standard deviation of azimuth spread values, it is straightforward to use (24) to predict the throughput degradation with increase in number of sectors. The variation of total capacity as a function of and $N$ can be observed in Fig. 12. It can be concluded that in the absence of angular spread, increase in number of sectors linearly increases the total capacity. However in the presence of angular spread, the increase in total capacity saturates after a certain number of sectors. This breakpoint on number of sectors is dependent on the angular spread. Note that (24) is derived with the assumption of ideal rectangular antenna patterns.

## X. Field trial of LTE-based Luneburg lens antenna system

### A. New Mexico Trial

The trial was performed in a rural area near a town called Truth or Consequences. We quantified the coverage and capacity of SuperCell in an area mostly composed of barren land.

Test Site Configuration:
- Spectra used: LTE TDD Band 41
- 6 Radios supporting 12 TDD cells

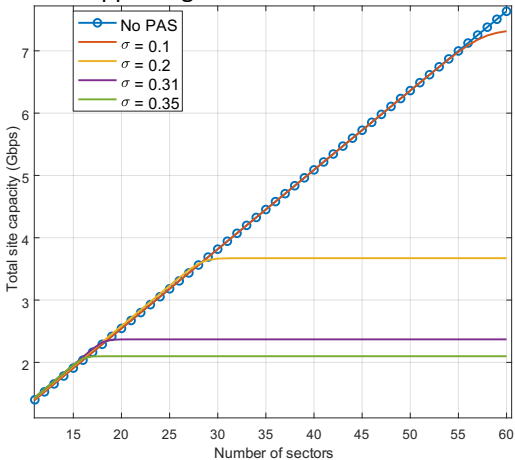

Fig. 12: Effect of azimuth spread on the capacity scaling with number of sectors, using (24, Appendix A) for a baseline CNR =20 $dB$

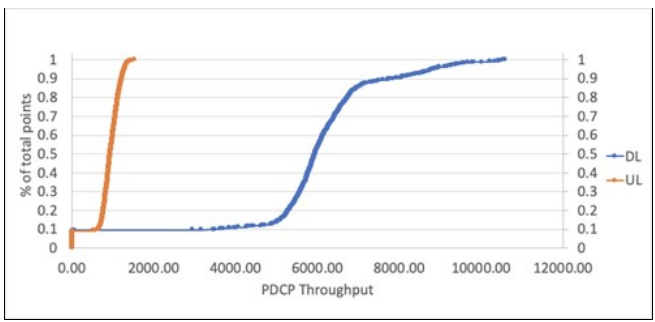

Fig. 13: Distribution of the DL and UL throughput values measured at 40km
SuperCell

- EPC-in-a-Box
- 1 Luneburg Lens Antenna 29 dBi gain
- EIRP 68dBm (according to the local transmit power limitations)
- Tower height: 10m on a hill with 250-300m higher elevation that the covered area

*1)    Performance Validation at 40km from SuperCell Tower:* The test was conducted under no-load conditions and focus was on determining coverage limits for the SuperCell. All devices were able to close the link (DL and Uplink) at a distance of 40km from the site. The RSRP measured was 116dBm with an SINR of 6dB. Average DL throughput seen on the fixed wireless access (FWA) device was 3.7 Mbps with an UL throughput of 1.1 Mbps.

For the test phone, the distribution of DL and UL throughput are shown in Fig. 13. With median values of 5.9kbps and 0.9kbps, respectively, DL throughput is about 7 times the UL throughput. This fact indicates a link imbalance. A partial remedy to the imbalance observed might be to use a tower mounted amplifier (TMA), which is a common practice in the industry to boost the uplink reception in rural sites particularly for high frequency bands.

*2)    Mobility Performance in Mid to Far Cell Range:* During the UE mobility drives, the test phone registered a single handoff failure, only, but there were a large number of radio link reestablishment failures observed in mid to far cell range. Detailed analysis of failure events revealed that the root cause is poor coverage, i.e. lack of any cell that is better than -120dBm, in areas which are in the "null" of the beams. Coverage prediction results were already pointing to this problem, however the predicted coverage gaps were showing up beyond 20 km, whereas in the field coverage gaps were experienced as close as 15 km from the tower. Potential mitigation techniques for this problem include:
- Incorporating the observed limitation as part of SuperCell design guidelines to avoid major roads covered from over 15 kms away
- Inter-locking design with the adjacent SuperCell, if any, so main beams from one cover the nulls of adjacent
- Adaptive sectorization/beam width

*3) Impact of Inter-Sector Interference:* Downlink: A test was performed to benchmark the impact of inter-sector interference on SINR and throughput. It was repeated twice at the same location, which is in the overlap zone of two adjacent sectors at the mid-cell range i.e. 14km from the site:
- With no load on serving or adjacent cells (baseline)
- 100% downlink load on adjacent cells A nearly 10 dB reduction in RS SINR (from 6dB median to -5dB) and a 60% reduction in DL throughput (from 11Mbps to 4.5Mbps) were observed under heavy interference from adjacent cells.

Uplink: A test was carried out to benchmark the impact of UL inter-sector interference at 3.7 km. The device under test (DUT) was placed in the boresight direction of its serving sector. The two interfering UEs were placed in the overlap areas with the two adjacent sectors. Each UE was connected to one of the three adjacent SuperCell sectors, hence, the UL receiver of the serving sector was receiving strong interference from the two UEs connected to adjacent sectors. UL throughput of DUT degraded from 17 Mbps baseline down to 10 Mbps under UL interference condition. So, a 40% reduction was observed.

## B. Mississippi Trial

In Mississippi, the trial was performed in the rural area near a town called Foxworth, where main vegetation is dense forest mostly containing very tall trees. We quantified the coverage and capacity of SuperCell and contrasted it against the typical macrocell coverage, which is only a few miles.

Test Site Configuration

- Spectra used: FDD, 10MHz in B25(B2) (1935-1945MHz)
- Spectra used: TDD, 20MHz in B41 (2530-2550MHz)
- Tower height: 178m
- 3 Luneburg Lens Antennas with 29 dBi gain
- 8 x B41 TDD Remote Radio Heads (RRHs) supporting 24 TDD sectors
- 12 x B2 FDD RRHs supporting 12 FDD sectors
- 2T2R and 2x2 MIMO
- TDD transmit power: 43dBm total (10W per port)

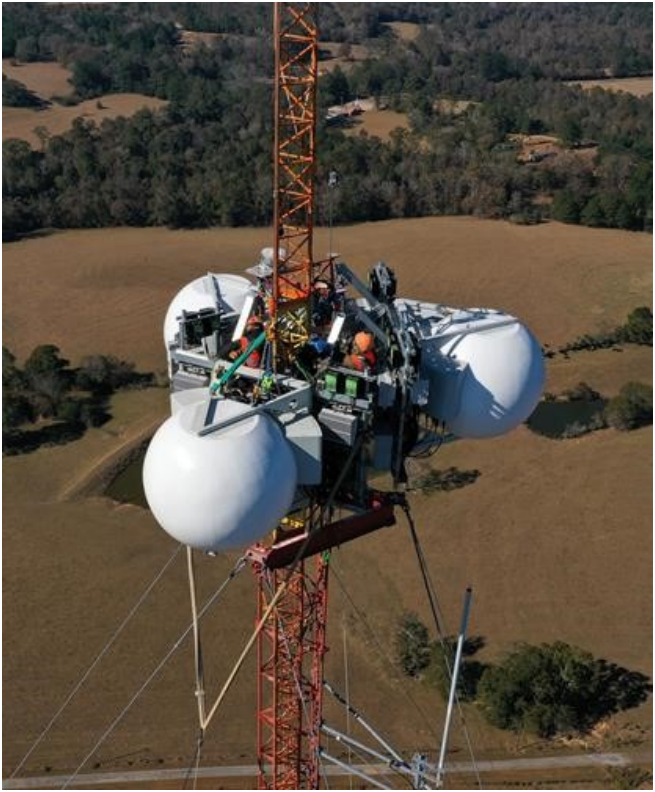

Fig. 14: A drone view of SuperCell base station at 180m tall tower

- 6dB RS boosting
- FDD transmit power: 46dBm total (20W per port)
- 6dB RS boosting

SuperCell had 12 FDD sectors oriented towards the West, and 24 TDD sectors that were oriented to the North, South and East. Antenna 1 and Antenna 3 transmit both TDD and FDD sectors, while Antenna 2 transmits TDD sectors, only. In this article, we will share the findings of FDD tests, only, due to space limitations.

*1) Brown-Field Deployment Test Results:* With 50% loading, the median SINR provided by the macrocell alone is 5.2dB. When SuperCell is turned on (also with 50% loading), it causes a significant drop in SINR: the median SINR reduces to 0.9dB. The SINR drop is especially large in the immediate vicinity of the macrocell since the macrocell signal was strong before SuperCell was turned on, and now both SuperCell and macrocell signals are strong there. For the near field areas, one could consider decommissioning the macrocell when SuperCell is activated. If macrocell is decommissioned, and only the SuperCell is active: the RSRP in the near field is actually ↕12dB better than what the macrocell can provide on its own. If macrocell is decommissioned, and only the SuperCell operates in the area, the median SINR drops from 5.2dB to 2.5dB.

*2) Mid-Field (10 km) Macrocell Test Results:* With 50% loading, the median SINR provided by the macrocell is 3dB. When SuperCell is turned on (also with 50% loading), it causes a mild drop in median SINR to 2.5dB, but considering that in coverage holes SINR measurements are not available, it can be stated that SINR has improved. The SINR improvement is considerably high for the cell edge users. If it was considered to decommission the macrocell when SuperCell is activated:

- the RSRP around the mid-field macrocell is still good, actually slightly better than what the macrocell provides

- the SINR around the mid-field macrocell is still OK, only 2dB less than what the macrocell provides

*3) Far-Field (16 km) Macrocell Test Results:* ICI is strong between SuperCell and Macrocell when they are both active. SuperCell, by itself, cannot serve the far field very well. An alternative solution is to split the available spectrum between SuperCell and macrocell (as 5MHz for SuperCell and 5MHz for macrocell). In that case, a good SINR with a median value of 5.1dB is obtained at the expense of using half the available bandwidth.

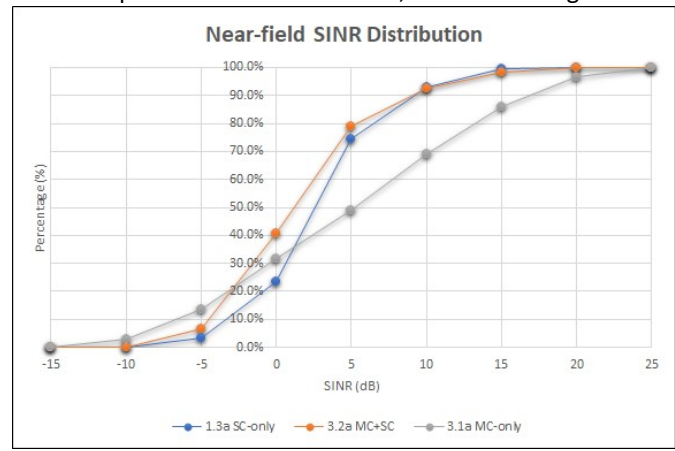

Fig. 15: Distribution of SINR values measured in the near-field in 3 different topologies

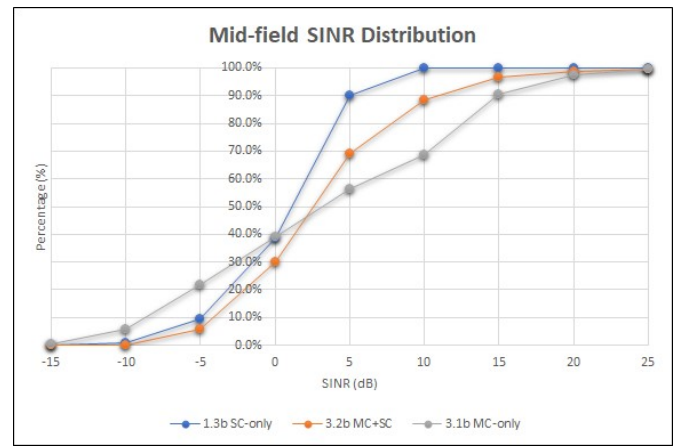

Fig. 16: Distribution of SINR values measured in the mid-field in 3 different topologies

## XI. Closing Remarks

In this paper, we proposed a novel cellular coverage and capacity enhancement solution called SuperCell. We identified and addressed several technical gaps in literature, in particular the lack of channel models to validate coverage and sectorization gains using such high-altitude platforms. Theoretical analysis and field measurement campaign was conducted to collect data in order to prove the feasibility of SuperCell. We also addressed antenna systems engineering challenges for such long-range elevated platform base stations in rural cellular connectivity. Following this, we also discuss a realworld type of cellular LTE deployment using Band 41 in two US cities in Mississippi and New Mexico. We addressed coexistence issues with legacy BTS in vicinity of a SuperCell site.

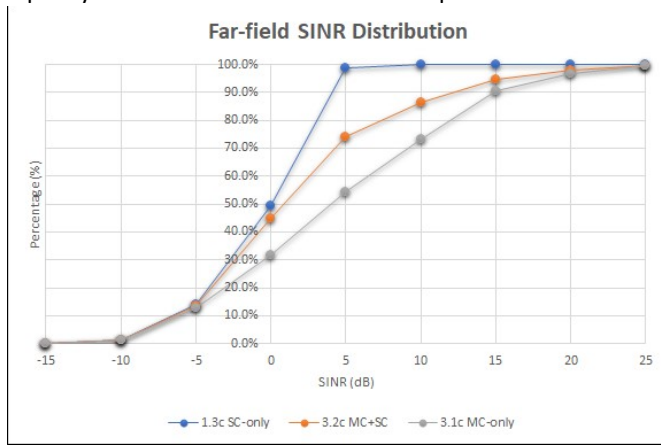

Fig. 17: Distribution of SINR values measured in the far-field in 3 different topologies

In this paper we show that it is possible to use commercially-available antenna systems with little engineering investment to realize a highly-efficient SuperCell, tailored for wide-area rural coverage, resulting in attractive coverage-cost proposition. The choice of antennas, and understanding of propagation characteristics, are key to proving its technoeconomic viability. Future research in identifying better class of system-level solutions that can further optimize coverage, capacity and cost based on a given techno-economic cost model.

## Acknowledgement

We would like to thank Sudhi Kandi, Afaq Arif, Rohan Ramakrishnan, Anthony Pham, Farbod Tabatabai, Anna Lindaseth, Anne Pak, Sonny Valino, Andy Cashion, Duane Drake and Bob Proctor from Facebook Connectivity for their immense contribution to custom software & hardware development along with accurate scheduling of various third-party vendor deliverables that enabled various field tests discussed in this paper.

We also extend our thanks to the support offered by academia. Prof. Arogyaswami Paulraj from Stanford University, California, Prof. Claude Osteges from UC Louvian, Belgium and Prof. Eric Michielssen from University of Michigan, Ann Arbor for their insightful technical guidance and in-depth review of this paper.

The authors would like to thank American Tower Corporation, C Spire and VoltServer for their partnering contributions towards Facebook Connectivity's Mississippi SuperCell trial. ATC for offering the use of a 180m tower, C Spire for offering Band 41 and Band 25 spectrum and VoltServer for offering their Digital Electricity power remoting solution for the trial respectively.

# Appendix A

## Capacity model: HOS & Inter-sector interference

### Analysis

When higher-order sectoring is employed to solve the huge capacity demands, there are two contributors to the inter-sector interference: antenna side lobe to/from neighboring sectors, and scattering caused by the clutter in the environment. We model these effects by simulated antenna radiation pattern and a two-ray model. The work of [39]–[41] show that the angular spread of the signal is modeled as a Laplacian or modified Gaussian power azimuth spectrum (PAS) for a variety of propagation environments.

### A. Relationship between PAS and Pathloss

Consider an uplink scenario: An omni directional Tx placed on the ground (approximately 2m height) and Rx on an elevated platform with a radiation pattern $R(\checkmark, )$, with $\checkmark$ and representing the azimuthal and elevation axis, respectively. The received signal power $P_r$ is given by

$$P_r = \int_{180}^{180} \int_{180}^{180} \Lambda(\theta, \phi) \mathscr{R}(\theta, \phi) d\theta d\phi \tag{8}$$

Assuming a two-ray model [15], where $T_1$ and $T_2$ are the resulting complex coeffcients from the two-ray model, and a Laplacian model for PAS with a standard deviation for direct path and $i$ for the reflected path, we obtain

$$P_r = T_1 \Lambda_{0,1} \int_{180}^{180} e^{\frac{\sqrt{2}|\theta|}{\sigma}} \mathscr{R}(\theta, 0) d\theta \\ + T_2 \Lambda_{0,2} \int_{180}^{180} e^{\frac{\sqrt{2}|\theta|}{\sigma_i}} \mathscr{R}(\theta, \phi_i) d\theta \tag{9}$$

Considering only the direct path i.e when there are no reflections, (9) boils down to

$$P_{r,dir} = \leftarrow_0 R(\checkmark_{dir, dir}) \tag{10}$$ where $\leftarrow_0$ includes the Tx power, Tx antenna gain and other cable losses. The pathloss can then be expressed as

$$PL = P_{r,dB,dir} \quad G_{rx,dB} \quad P_{t,dBm} \quad C_{l,dBm} \quad G_{tx,dB} \tag{11}$$

We can now define a reflection fade margin as

$$M_{ref,dB} = P_{r,dB} \quad P_{r,dB,dir} \tag{12}$$

The resulting pathloss in the presence of multipath can now be expressed as

$$PL = P_{r,dB} \quad M_{ref,dB} \quad G_{rx,dB} \quad P_{t,dBm} \\ C_{l,dBm} \quad G_{tx,dB} \tag{13}$$

Note that the margin $M_{ref,dB}$ is not deterministic as it depends on the time dependent variation in the environment and the location of reflecting surfaces relative to the transmitter. A well calibrated 3D ray tracing simulation model is required to determine the standard deviation of $M_{ref,dB}$ for a given type of environment.

### B. Relationship between PAS and C/I

Considering the PAS for azimuth axis only, desired signal power $D_p$ is given as

$$D_p = P_t \int_0^{360} \Lambda_d(\theta) \mathscr{R}(\theta) d\theta \tag{14}$$

where $\leftarrow_d$ is the PAS for the desired UE location. Assuming $N$ interferers, the total interference power, $I_p$ with their relative azimuthal location with respect to the desired Tx location as $\checkmark_n$, is given by

$$I_p = \sum_{n=1}^{N} P_{t,n} \int_0^{360} \Lambda_{i,n}(\theta) \mathscr{R}(\theta \quad \theta_n) d\theta \tag{15}$$

where $P_t$ represents the power transmitted by the $n^{th}$ UE and $\leftarrow_{i,n}$ represents the PAS for the $n^{th}$ interferer. C/I ratio can now be expressed as

$$C/I = \frac{D_p}{I_p} \\ = \frac{P_t \int_0^{360} \Lambda_d(\theta) \mathscr{R}(\theta) d\theta}{\sum_{n=1}^{N} P_{t,n} \int_0^{360} \Lambda_{i,n}(\theta) \mathscr{R}(\theta \quad \theta_n) d\theta} \tag{16}$$

If channel reciprocity is assumed i.e the multipath components are identical in uplink and downlink, it is straightforward to prove that (16) can be applied for a downlink scenario as well.

### C. Uniform sectorization capacity model

Consider a uniform HOS model constructed using ideal saw-tooth antenna patterns. For an N-sector system, the width of the pass-band of the $n^{th}$ sector is given by $W_{pass} = \frac{180}{N}$

Using the Laplacian model for PAS, the desired power $D_p$ and the interference power $I_p$ for $0^{th}$ sector is given by

$$D_p = \Lambda_0 \int_{180}^{180} e^{\frac{\sqrt{2}|\theta|}{\sigma}} \mathscr{R}(\theta) d\theta \tag{17}$$

$$I_p = \sum_{n=1}^{N} \Lambda_0 \int_{180}^{180} e^{\frac{\sqrt{2}|\theta|}{\sigma}} \mathscr{R}(\theta \quad nW_{pass}) d\theta \tag{18}$$

Note that (17) and (18) assume that all transmitters are equally scattered and located in the center of their desired sectors. Using the definition of $R(\checkmark)$, the integrals in (17) and (18) can be solved and simplified as

$$D_p = \text{p2} \leftarrow_0 \left( 1 \quad e^{\frac{\sqrt{2}W_{pass}}{\sigma}} \right) \tag{19}$$

$$I_p = \text{p2} \leftarrow_0 \sqrt{} e \qquad e^{\frac{\sqrt{2}NW_{pass}}{}} \qquad e^{\frac{\sqrt{2}W_{pass}}{}} \qquad \blacklozenge \tag{20}$$

Defining $K_o$ as the Boltzmann constant, $B$ as the bandwidth, and $T$ as the temperature, the thermal noise power can be expressed as

$$N_p = K_o TB \tag{21}$$

The carrier-to-interference-noise ratio can now be defined as

$$\text{CINR} = \frac{D_p}{I_p + N_p}$$

$$= \frac{\sqrt{2}\sigma\Lambda_0\left(1 - e^{-\frac{\sqrt{2}W_{pass}}{\sigma}}\right)}{\sqrt{2}\sigma\Lambda_0\left(e^{-\frac{\sqrt{2}NW_{pass}}{\sigma}} - e^{-\frac{\sqrt{2}W_{pass}}{\sigma}}\right) + N_p} \qquad (22)$$

The carrier-to-noise ratio in the absence of any interfering transmitter is given by

$$\text{CNR} = \frac{\sqrt{2}\sigma\Lambda_0\left(1 - e^{-\frac{\sqrt{2}W_{pass}}{\sigma}}\right)}{N_p} \qquad (23)$$

From (22) and (23) it can be inferred that as → 0 i.e towards lower angular spread, CINR → CNR. The resulting of total capacity (for all N sectors) can be analyzed using the Shannon limit theorem as shown in (24). Once there is an prediction of azimuth spread from the measurement results, (24) can be revisited to estimate the achievable capacity for a given number of sectors.

$$C_{tot} = 2NB\log_2(1 + \text{CINR}) \qquad (24)$$